\newcommand{\be}{\begin{eqnarray}}
\newcommand{\ee}{\end{eqnarray}}

\def\1{{\rm 1\mskip-4.5mu l} }

\documentclass[12pt]{article}

\usepackage{amsmath,amssymb}
\usepackage{graphicx}
\usepackage{cite}
\usepackage{bm}
\begin{document}
\renewcommand{\thefootnote}{\fnsymbol{footnote}}

\hfill AEI-2008-058

\vskip 15mm

\begin{center}

{\Large Exceptional points in quantum and classical dynamics}

\vskip 4ex

A.V. \textsc{Smilga}

\vspace{.5cm}

SUBATECH, Universit\'e de
Nantes,  4 rue Alfred Kastler, BP 20722, Nantes  44307, France
\footnote{On leave of absence from ITEP, Moscow, Russia.}

$\texttt{smilga@subatech.in2p3.fr}$
\end{center}

\vskip 5ex

\begin{abstract}
\noindent  We notice that, when a 
quantum system involves exceptional points,
i.e. the special values of parameters where the Hamiltonian loses its 
self-adjointness and acquires the Jordan block structure, the corresponding 
classical system also exhibits a singular behaviour associated with 
restructuring of classical trajectories. The system with the crypto-Hermitian
Hamiltonian $H = (p^2+z^2)/2 -igz^5$ and hyper-elliptic classical dynamics is 
studied in details. Analogies with supersymmetric Yang-Mills dynamics are elucidated.
\end{abstract}

\renewcommand{\thefootnote}{\arabic{footnote}}
\setcounter{footnote}0
\setcounter{page}{1}

\section{Introduction}
A sufficient condition for the spectrum of a Hamiltonian to be real is its 
 self-adjointness, $H^\dagger = H$. However, there exists a rich class of
 systems whose Hamiltonian is not manifestly Hermitian, but the spectrum
 is still real. Probably, the simplest such example is the matrix Hamiltonian
 \be
\label{2by2}
H = \left( \begin{array}{cc} 1 & 1 \\ 0 & 2 \end{array} \right)
 \ee
with two real eigenvalues $\lambda_1 = 1$ and $\lambda_2 = 2$. Such systems 
have been intermittenly discussed  since mid-seventies \cite{brmog}, 
but the interest to this problem was considerably boosted after the paper \cite{BB} appeared, 
 where the systems with {\it complex} potentials $V(x) = x^2(ix)^\epsilon$ were studied
and it was shown  that the spectrum of the corresponding Hamiltonians is real.   

Generically, a Hamiltonian involving only real eigenvalues can be transformed in a manifestly
Hermitian form by a similarity transformation \cite{Mostafa}, 
$H \to RHR^{-1}$. If $H$ is not manifestly Hermitian, $R$ is not unitary. 
This amounts to modifying the Hilbert space measure so that the 
probability 
$P = \langle \psi |M|\psi \rangle$ defined with the new measure $M$ is conserved and the theory
is unitary. 
\footnote{The explicit forms of $R$ and $M$ for the Hamiltonian (\ref{2by2}) are
\be
\label{RM} 
R =   \ \left( \begin{array}{cc} 1 & -1  \\  0 & 1 \end{array} \right) \ \ \ \ \ \ \ \ \ \ \ \ ,\ \ \ \ 
M = R^T R \ =\ \left( \begin{array}{cc} 1 & -1  \\  -1 & 2 \end{array} \right)
 \ee
}  One can call a Hamiltonian of this type {\it crypto-Hermitian} (Hermitian in disguise) 
\cite{Zee,Ivanov}.

However, not all Hamiltonians with real eigenvalues are crypto-Hermitian. In some cases, 
a Hamiltonian cannot be rendered Hermitian (and eventually diagonalized) by a similarity transformation.
The simplest example is a $2\times 2$ matrix representing a Jordan block, $H = (\begin{array}{cc}1&1\\0&1
\end{array} )$. The {\it exceptional points} when this happens 
are associated with degeneracy of eigenvalues \cite{Heiss}. Exceptional points have measure zero in the 
space of parameters.

 There are exceptional points involving multidimensional Jordan blocks. For the higher-derivative 
  Pais-Uhlenbeck oscillator \cite{PU} at equal frequencies, the dimension of emerging Jordan blocks is even infinite
\cite{duhi,comment} . A more typical kind of exceptional point
 is when only a couple 
of eigenvalues coalesce. Changing  the parameters beyond the exceptional point, a pair of complex
eigenvalues that are conjugate to each other appears.
\footnote{Again, the simplest example is the matrix 
$H = (\begin{array}{cc} 1&1 \\ \alpha & 1 \end{array} )$. 
When $\alpha$ is small
and positive, we have a pair of close real eigenvalues. When $\alpha$ is small and negative, there is a pair of 
complex conjugate eigenvalues. $\alpha = 0$ is the exceptional point. } 

A numerical study performed in Ref.\cite{BB} displayed an infinite number of 
exceptional points in the parameter $\epsilon$ for the Hamiltonian
 \be
 \label{Heps} 
H = p^2 + z^2(iz)^\epsilon\ .
 \ee
 The exceptional points lie in the interval
$\epsilon \in (-1,0)$. The leftmost exceptional point is 
$\epsilon_* \approx -.578$. When $\epsilon < \epsilon_*$, the spectrum of the Hamiltonian involves
only one real eigenvalue (the ground state), with all other eigenvalues coming in complex conjugate pairs.
At $\epsilon = \epsilon_* $, one of such pairs coalecses and, at still larger $\epsilon$, 
is transformed into a pair of close real eigenvalues. Then comes the turn of the second pair, etc. 
The infinite number
of exceptional points  has an accumulation point at $\epsilon =0$ 
\cite{DorTat}.
 At $\epsilon > 0$, complex eigenvalues disappear and      
the Hamiltonian is crypto-Hermitian. 

A similar phenomenon was observed in Ref.\cite{crypto} for the Hamiltonian
  \be
\label{Hx5}
H \ =\ \frac{p^2+z^2}2 -igz^5 \ .   
  \ee
There are {\it two} different spectral problems corresponding 
to this Hamiltonian (see  Refs.\cite{DoreyShin,crypto} for 
detailed explanations). One of the problems is defined in  the Stokes wedges
 \be
\label{wedges0}
- \frac {\pi}{14} < &{\rm Arg}(z)& <   \frac {3\pi}{14}\ , \nonumber \\
\frac {11\pi}{14} < &{\rm Arg}(z)& <  \frac {15\pi}{14} 
  \ee 
including the 
real  axis. It turns out that the spectrum is real there for all values of
$g$. But for another spectral problem defined in the wedges
 \be
\label{wedges}
- \frac {5\pi}{14} < &{\rm Arg}(z)& <  - \frac {\pi}{14}\ , \nonumber \\
\frac {15\pi}{14} < &{\rm Arg}(z)& <  \frac {19\pi}{14} 
  \ee 
in the complex $z$ plane, the situation is different. The spectrum is real 
(and the Hamiltonian is crypto-Hermitian) for large enough $g$, but the 
system involves an infinite number of exceptional points at small $g$. If 
$g < g_* \approx 0.037$, a pair of complex eigenvalues appears. If $g < g_{**}
 \approx 0.007$, 
there are at least two such pairs, etc.

 A question that can be asked is whether the presence of 
these {\it quantum} exceptional points
displays themselves in some way also in the dynamics of the corresponding
 {\it classical} systems ? A partial answer to this question was obtained in
Ref.\cite{Bendnew}  where a series of critical values of the parameter 
$\epsilon$,  where the pattern of classical trajectories in the problem (\ref{Heps}) is changed, 
was found. The values of the classical exceptional points do not coincide with the values
of the quantum exceptional points. One can only say that the {\it presence} of the former is associated
with the presence of the latter.    

The problem (\ref{Heps}) is rather complicated, however. The Riemann surface for the potential
$\sim z^{2+\epsilon}$ has generically an infinite number of sheets, the classical trajectories can visit a lot
of the sheets and look complicated. That is why we decided to study this question for the system
(\ref{Hx5}), which is much simpler. The classical trajectories represent in this case hyper-elliptic functions
 known to mathematicians \cite{hyperell}.  

The result is the following. There are no classical exceptional points for the system (\ref{Hx5}) with
a positive energy. When the energy is negative, there is  a {\it single} classical exceptional point 
 \be
\label{clasexc}
g_*^{\rm class} \ =\ \frac 15 \left(-  \frac 3{10E} \right)^{3/2} \approx \ 
\frac {0.0329}{|E|^{3/2}} \ .
 \ee
As was mentioned above, a classical exceptional point is the point where the pattern  of
the classical trajectories changes. For the system (\ref{Hx5}), the reason for this change is
very simple: it happens that at $g = g_*^{\rm class}$ two of the five {\it turning points}, i.e. two
of the five solutions to the equation
 \be
\label{eqnx5x2}
V(x) = \ \frac{z^2}2 - igz^5 = E \ < 0
 \ee
coalesce. This phenomenon has a lot in common with the so called {\it Argyres-Douglas} phenomenon
 in   supersymmetric Yang-Mills theory \cite{AD}). We will dwell upon this issue  in Sect. 4. 

 Before discussing the dynamics of the system (\ref{Hx5}), we want to make few 
comments on the classical dynamics of a simpler system
 \be
 \label{Hx3}
H \ =\ \frac {p^2+\omega^2 z^2}2 + igz^3 \ ,
 \ee
 where exceptional points do not appear either at the classical or at the quantum level.

\section{Complex cubic potential and elliptic dynamics.}
 Consider the system (\ref{Hx3}). Complex classical trajectories are the solutions to the equation
 \be
\label{z3E}
\frac {\dot{z}^2}2 + \frac {\omega^2 z^2 }2 + igz^3 = E\ .
 \ee
The solutions to this equation with real positive energies (especially, in the case $\omega = 0$) were
 found numerically in
Ref.\cite{BB}. The solutions with real negative energies were found in Ref.\cite{crypto}. A simple remark that
we want to make here is that the solution to Eq.(\ref{z3E}) has a name - it is the Weierstrass function. Let
for simplicity set $\omega = 0, g=1$.
The equation is brought into the canonical Weierstrass form \cite{ellipt},
 \be
\label{canform}
\dot{y}^2 = 4y^3 - g_2y - g_3\ ,
 \ee 
with $g_2 = 0, g_3 = E/2$, by setting $z = 2iy$.

The Weierstrass function $z = P(t, g_2, g_3)$ is an elliptic double-periodic complex function. 
Generically, the periods are complex, but in our case one of the periods is real.  For positive energies, 
 \be
\label{T1}
T_1 = 5 \sqrt{\frac \pi 6} \frac {\Gamma(4/3)}{\Gamma(11/6)}E^{-1/6} \approx 3.434 E^{-1/6}\ .
  \ee
Another period is still complex, $T_2 = T_1 e^{i\pi/3}$. 
$T_2$ is obtained from $T_1$ by rotation in the complex $E$ plane, $E \to Ee^{-2i\pi}$.
Any linear combination of $T_1$ and $T_2$ with integer coefficients is also a period. A purely imaginary
combination $i\tilde{T} = 2T_2 - T_1 = i\sqrt{3}T_1$ represents a particular interest. 
 The physical interpretation of the periods $T_1$ and $\tilde{T}$ is clear \cite{BB,crypto}. 
The periodicity with respect to the real
time shift $t \to t + T_1$ is a usual physical periodicity of the trajectories. For all trajectories, the period
is the same. The imaginary time shift transforms one trajectory into another. 
\footnote{This transformation can be 
interpreted as a gauge transformation \cite{crypto}, but it is not a point of interest for us now.}

One can also be interested with the solutions to Eq.(\ref{z3E}) at complex $E$. It is still Weierstrass function, 
but all its periods are now complex. This means that the solution is  not
 periodic in real physical time. Such solutions were studied numerically in Ref.\cite{BBH}.

The family of all trajectories contains two distinguished members that can be called {\it stem} trajectories.
One of the stem trajectories connects two turning points (solutions to the equation $V(z) = iz^3 = E$)
in the lower half-plane, $z_1 = E^{1/3} e^{-5i\pi/6}$ and  $z_2 = E^{1/3} e^{-i\pi/6}$. Another stem trajectory
goes from the turning point $z_3 = iE^{1/3}$ to infinity. The shift $t \to t + i\tilde{T}/4$ transforms one stem
trajectory into another. The shift $t \to t + i\tilde{T}/2$ transforms the function $z = 2iP(t,0,E/2)$ to
$z = 2iP(-t,0,E/2)$ and the shift by $i\tilde{T}$ leaves the Weierstrass function intact. 
This is all illustrated in Figs.1,2.

\begin{figure}[h]
   \begin{center}
 \includegraphics[width=2.0in]{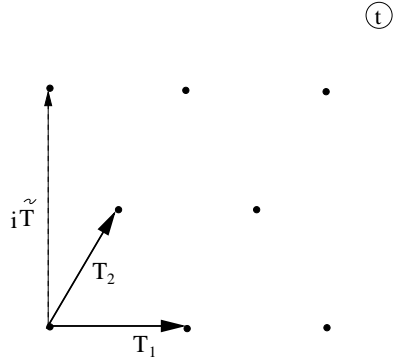}
    \end{center}
\caption{Periods of the Weierstrass function.}
\label{periody}
\end{figure}

 \begin{figure}[h]
   \begin{center}
 \includegraphics[width=4.0in]{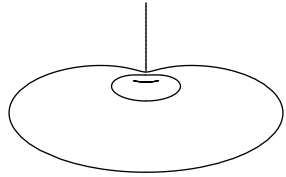}
    \end{center}
\vspace{-10cm}
\caption{The trajectories $ z = 2iP(t+ia, 0,1/2)$,  \ $t \in (0,T_1)$.
 The vertical line ($a =0$) is the stem trajectory connecting the
 branching points $z = i$ and $ z = \infty$. The ``smile'' ($a=\tilde{T}/4$) is the stem trajectory connecting
the branching points $z= e^{-5i\pi/6}$ and $z = e^{-i\pi/6}$. The trajectories with intermediate values $a = \tilde{T}/16$
and $a = \tilde{T}/8$ are also plotted.} 
\label{Weierstrass}
\end{figure}

Translating these observations into a standard mathematical language, one can say that the Weierstrass function 
describes the motion over the Riemann surface of the function $\sqrt{E-iz^3}$. The turning points and infinity 
are nothing but the branching points of this function. This Riemann surface has the topology of torus. 
Two periods $T_{1,2}$ corresponds to two cycles of this torus.  
 
 One can also plot the imaginary time trajectories $2iP(a+i\tilde{t}, 0 ,1/2)$, \ $\tilde{t} \in (0, \tilde{T})$
and observe by inspecting Eq.(\ref{canform}) that they {\it coincide} up to a sign with the real time
trajectories $2iP(\tilde{t} - ia, 0, -1/2)$ having negative energies and studied in Ref.\cite{crypto}.

\section{The potential $V(z) = z^2/2  - igz^5$ and hyper-elliptic dynamics.}

  Classical complex trajectories for the Hamiltonian (\ref{Hx5}) have much in common with the elliptic
 trajectories of the previous section. They are hyper-elliptic trajectories decribing the motion over the Riemann
surface of the function $\sqrt{E - V(z)}$. This Riemann surface has genus 2 and two pairs of cycles.
The latter implies that the equation
 \be
 \label{eqz5}
\frac {\dot{z}^2}2 + V(z) = E 
 \ee
has now  different types of solutions stemming from the trajectories that connect three different pairs of 
 branching points of the function $\sqrt{E-V(z)}$.
 Let first the energy be positive and $V(z) = -iz^5$ (without the quadratic term). The branching points are in the vertices
of the pentagon and at infinity. Three different stem trajectories are depicted
in Fig.3. The corresponding periods are \cite{BB,crypto}

\begin{figure}[h]
   \begin{center}
 \includegraphics[width=1.5in]{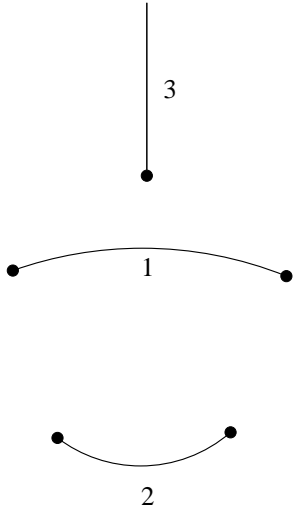}
    \end{center}
\caption{Stem trajectories for the potential $-iz^5$}
\label{pentagon}
\end{figure}

\be
\label{Reper}
T_1 &=& \frac 75 \sqrt{2\pi} \cos \frac {\pi}{10} \frac {\Gamma(6/5)}{\Gamma(17/10)} E^{-3/10}  \nonumber \\
T_2 &=& \frac 75 \sqrt{2\pi} \cos \frac {3\pi}{10} \frac {\Gamma(6/5)}{\Gamma(17/10)} E^{-3/10} .
 \ee
 and $T_3 = T_1 - T_2$. 

Similarly to what we had for the cubic potential, the solutions to Eq. (\ref{eqz5}) are also periodic
with respect  to imaginary time shifts,

 \be
\label{Imper}
i\tilde{T}_1 &=&  i \frac 75 \sqrt{2\pi} \left( 1 + 2 \sin \frac {3\pi}{10}
+ \sin \frac \pi{10} \right) \frac {\Gamma(6/5)}{\Gamma(17/10)} E^{-3/10} \nonumber \\ 
i\tilde{T}_2 &=&  i \frac 75 \sqrt{2\pi} \left( 1 +  \sin \frac {3\pi}{10} \right)
 \frac {\Gamma(6/5)}{\Gamma(17/10)} E^{-3/10} \nonumber \\ 
i\tilde{T}_3 &=&  i \frac 75 \sqrt{2\pi} \left( \sin \frac {3\pi}{10}
+ \sin \frac \pi{10} \right) \frac {\Gamma(6/5)}{\Gamma(17/10)} E^{-3/10} 
 \ee

The periods (\ref{Reper},\ref{Imper}) are all interrelated under rotations in the complex $E$ plane (the monodromies).
\footnote{Cf. Proposition 7 in Ref.\cite{Fedor}.}
 For
example,
 \be
\label{relper}
T_1(Ee^{-2i\pi}) \ =\ \frac {i\tilde{T_2} - T_2}2, \ \ \ \ \ \ T_2(Ee^{-2i\pi}) \ =\ \frac {i\tilde{T_3} - T_3}2
  \ee

This leads to three families of trajectories with positive energies and three 
families of trajectories with negative energies. 
\footnote{To be more precise, the third family is degenerate in this case and  consists in only one member: 
the stem trajectory starting at $z=iE^{1/5}$ and going away to $z = i\infty$ in finite time, $t_{\rm escape} = T_3/2$. }

Let us switch on the quadratic term now and investigate how the pattern of trajectories
is changed when changing $g$. Let first the energy be positive.  
Note that for all $g$, Eq.(\ref{eqnx5x2}) still has 5 distinct roots (see Fig.3).


\begin{figure}[h]
   \begin{center}
 \includegraphics[width=8in]{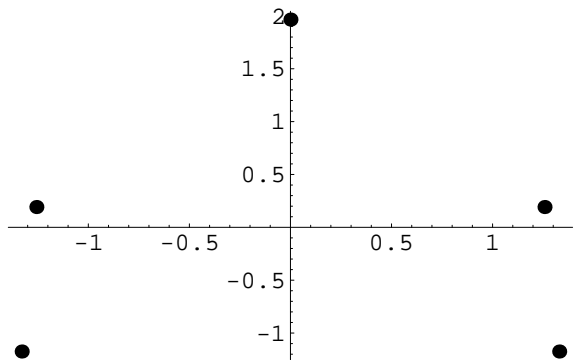}
    \end{center}
\vspace{-19cm}
\caption{Turning points. $E = 1,\ g=0.1$}
\label{kornipos}
\end{figure} 

We see that the turning points are still symmetric with respect to the imaginary axis. There are still three 
families of trajectories  stemming from the trajectories that have
the same qualitative pattern as in Fig. \ref{pentagon}. Nothing changes essentially when $g$ is changed.

\begin{figure}[h]
   \begin{tabular}{c@{\hspace{-4cm}}c}
 \hspace{1cm} \includegraphics[width=5in]{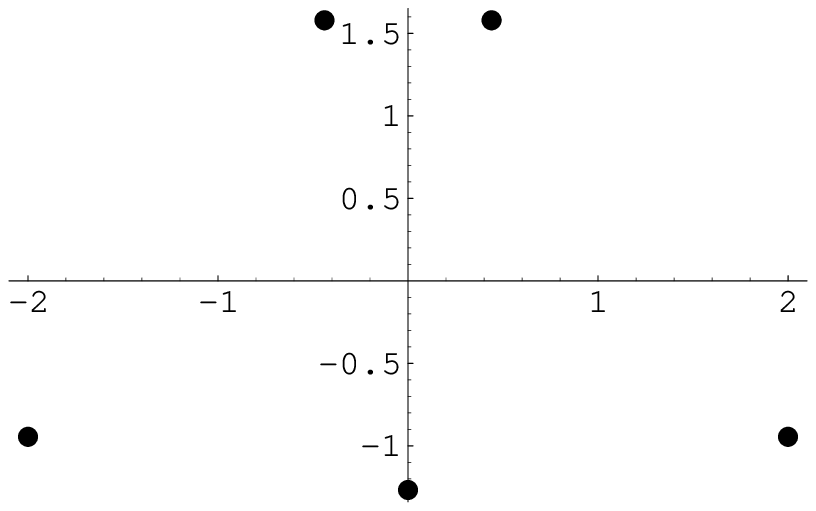} & \includegraphics[width=5in]{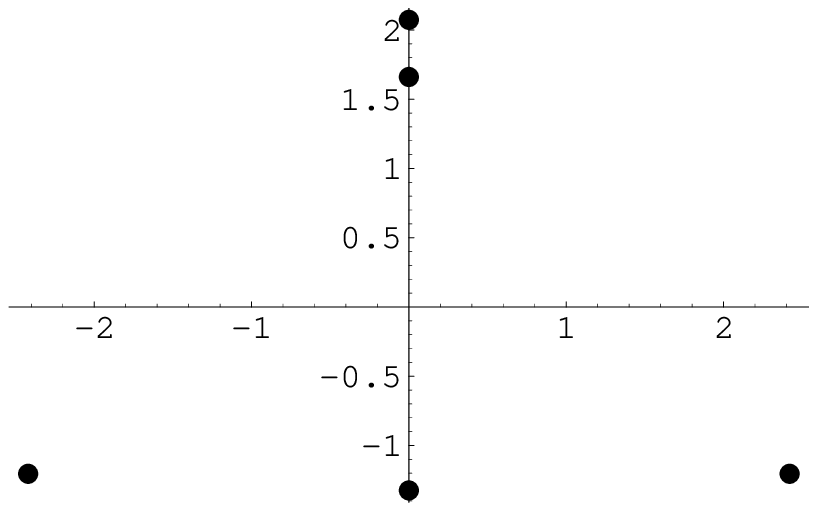}
\\ \vspace{-11cm} 
\\
\hspace{-4cm} a) & \hspace{-4cm} b)
    \end{tabular}
\caption{Turning points. $E = -1$; {\it a)} $g=0.06$, \ {\it b)} $g = 0.03$.}
\label{kornineg}
\end{figure}

When $E < 0$, the situation is different. There is a point (\ref{clasexc}) where two of the roots of Eq.(\ref{eqnx5x2})
 coalecse. The root
patterns slightly above the exceptional  point $g = g^*$ and slightly below it is shown in Fig. \ref{kornineg}.


\begin{figure}[h]
   \begin{tabular}{c@{\hspace{-5cm}}c}
 \hspace{1cm} \includegraphics[width=5in]{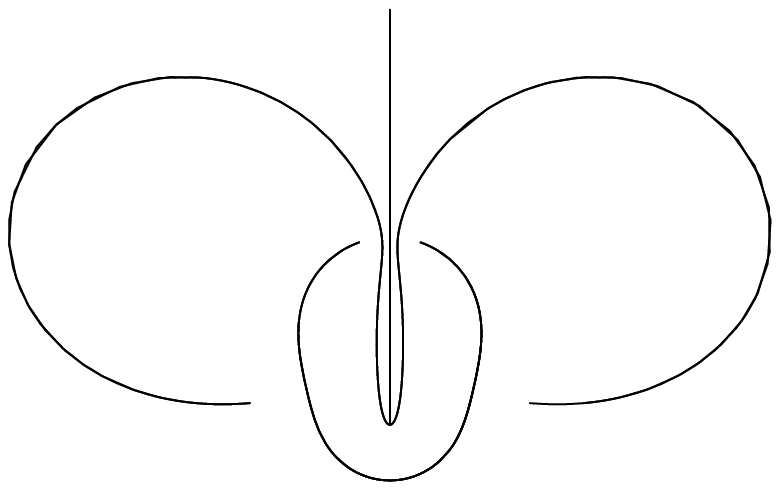} & \includegraphics[width=5in]{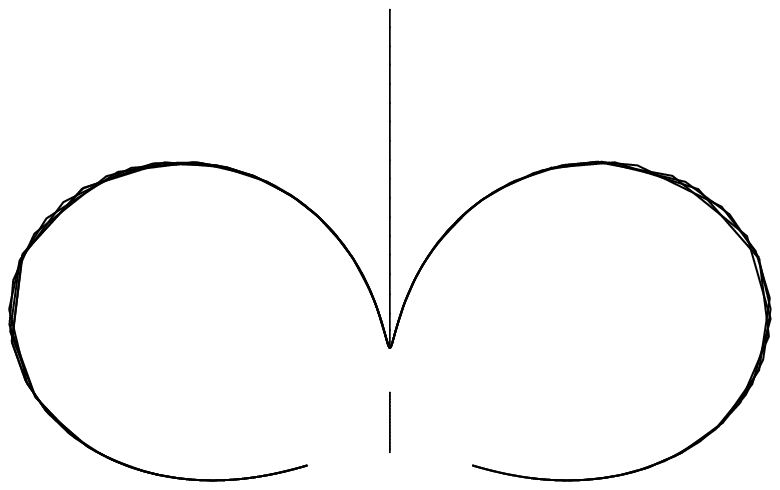}
\\ \vspace{-11cm} 
\\
\hspace{-4cm} a) & \hspace{-5cm} b)
    \end{tabular}
\caption{Stem trajectories. $E = -1$; {\it a)} $g=0.06$, \ {\it b)} $g = 0.01$. The 
 small gaps between
the vertical lines going to $i\infty$ and the curves below are  indistinguishable. }
\label{stemneg}
\end{figure}

When $g > g^*$, there are three families of orbits. Their stem trajectories are shown in Fig.\ref{stemneg}a. 
When $g < g*$, the pattern of the stem trajectories is different (see Fig. \ref{stemneg}b). One can notice that
the stem trajectory connecting the low pair of the turning points with Re$(x) \neq 0$, 
exists at both $g > g^*$ and $g < g^*$, but 
its  form is different. When $ g \to g^*+i0$, the trajectory crosses itself, after which the ``appendix''
going around the low turning point at the imaginary axis disappears.  
 

\section{Analogies with SYM.}
As was already mentioned, the restructuring of classical trajectories observed in this paper has
 much in common with the Argyres-Douglas phenomenon that occurs in SYM theory. Let us briefly explain this point here 
(see Ref. \cite{lazha} for more details).

${\cal N} =2$ supersymmetric gauge theories are known to include 
vacuum moduli spaces: valleys of degenerate classical vacua 
parametrized by vacuum expectation values of scalar  fields \cite{SW}.
These valleys have some distinguished singular points where the mass
of certain objects present in the theory (the monopoles and dyons) 
vanishes. The original AD observation was that, in some theories and 
for certain values of parameters,
the moduli space singularities might coalesce. 

An analogy between this phenomenon and the classical exceptional point (\ref{clasexc})
exists, but is not so manifest. It becomes much more clear for the theories where
${\cal N} =2$ is broken down to ${\cal N} =1$. The continuous degeneracy
of the vacuum valleys is lifted in this case, and we are left with a finite
number of different classical vacua. These vacua can be separated by domain walls, the classical
solutions to the equations of motion with proper boundary conditions. Mathematically,
domain walls interpolating between the classical vacua play exactly the same role 
as the stem trajectories interpolating between the turning points in the QM models
discussed above. 

The point is that, in some ${\cal N} =1$ models, the parameters can be chosen such 
that the classical vacua coalesce. This phenomenon is akin to
the original AD phenomenon (merging of singularities in vacuum moduli spaces). One can
call it the AD phenomenon of the second kind. It was discovered first 
for a somewhat exotic model based on the $G_2$
gauge group and involving besides ${\cal N} =1$ gauge supermultiplet three different
chiral multiplets $S^{i=1,2,3}_{\alpha = 1,\ldots,7}$ in the fundamental representation of
$G_2$ \cite{G2}. The superpotential of the model is
 \be
\label{WG2}
{\cal }W \ =\ -\frac m2 (S^i_\alpha )^2 - \frac \lambda 6 e_{ijk} f^{\alpha \beta \gamma} S_\alpha^i S_\beta^j S_\gamma^k \ ,
 \ee
where $f^{\alpha \beta \gamma}$ is the invariant antisymmetric tensor of $G_2$
\footnote{Note that $G_2$ can be defined as a subgroup of $SO(7)$ leaving invariant
the form $f^{\alpha \beta \gamma} A_\alpha B_\beta C_\gamma$ for any set of three different
7-vectors $A,B,C$.}.
 To analyze vacuum structure, one has to add to the superpotential (\ref{WG2}) a nonperturbative instanton--generated
superpotential of the Affleck-Dine-Seiberg type \cite{ADS}. In this case, it has the form
 \be
\label{ADS}
{\cal W}^{\rm inst} \ =\ \frac {\Lambda^9}{{\cal B}^2 - \det{\cal M}} \ ,
 \ee  
where ${\cal B}$ and ${\cal M}$ are the gauge-invariant moduli, 
 \be
\label{MB}
{\cal M}^{ij} =\ S^i_\alpha S^j_\alpha , \ \ \ \ {\cal B} = \frac 16 \epsilon_{ijk}  S_\alpha^i S_\beta^j S_\gamma^k 
 \ee 
 and $\Lambda$ is the confinement scale.

Adding the superpotentials  (\ref{WG2}) and (\ref{ADS}), we obtain a theory with  two dimensionless parameters: 
the Yukawa coupling $\lambda$ and the ratio  mass $m/\Lambda$.
 The equation determining the 
 chirally asymmetric classical vacua
\footnote{We leave aside a controversial issue of possible existence of the extra chirally symmetric
vacuum \cite{KS}.} is of the 6-th order,
 \be
\label{equ}
\frac {\partial {\cal W} }{\partial ({\rm moduli}) } = 0 \ \longrightarrow \ 
mu^4 \left( 1 - \frac {\lambda^2 u}{m^2}  \right)^2 =  \Lambda^9 \  ,
 \ee
where $u = \frac 13 {\rm Tr} \,{\cal M} $. Generically, there are 6 distinct solutions.  
 However, when   $\lambda^2 \approx  .385\, m^2/\Lambda^2 $,  two of these vacua coalesce.

The simplest model where the AD phenomenon of the second kind occurs is based on $SU(2)$ gauge group
and involves a massive adjoint chiral multiplet and a pair of fundamental chiral multiplets.
This model obtained by a deformation of ${\cal N} =2$ supersymmetric QCD (the model involving besides
the gauge ${\cal N} =2$ multiplet also a matter hypermultiplet) and analyzed in Ref.\cite{GVY}
involves generically 3 vacua, but, for some values of masses and Yukawa couplings, two of them coalesce. 

When the vacua merge, the domain wall interpolating between them disappears. The analogy with the classical exceptional
point (\ref{clasexc}) is obvious.

\vspace{1mm}

I aknowledge numerous illuminating discussions with A. Gorsky and warm hospitality at AEI institute in Golm where this 
work was finished.

\end{document}